\documentstyle[prl,aps,epsf,twocolumn]{revtex}

\begin{document}

\title
{Color Molecular-Dynamics for High Density Matter}

\author{Toshiki Maruyama$^{(1)}$ and Tetsuo Hatsuda$^{(2)}$}
\address{$^{(1)}$ Advanced Science Research Center, 
 Japan Atomic Energy Research Institute, Tokai, Ibaraki 319-1195,  Japan}
\address{$^{(2)}$  Physics Department, Kyoto University, Kyoto 606-8502, Japan}

\date{\today}

\maketitle

\begin{abstract}
We propose a microscopic simulation for quark many-body system 
based on molecular dynamics.
Using color confinement and  one-gluon exchange potentials together with
  the meson exchange potentials between quarks, 
we construct  nucleons and
 nuclear/quark matter.
Statistical feature and the dynamical change between 
confinement and deconfinement phases are studied with this
molecular dynamics simulation.
\end{abstract}


At high baryon density, the nuclear matter is believed to
undergo a phase transition to the quark matter because of the
color Debye screening and the
asymptotic freedom in quantum chromodynamics (QCD) \cite{CP}.
In qualitative estimates using the 
 Bag model \cite{Baym} as well as
 the strong coupling lattice QCD \cite{Kawa}
 predict a first order transition at baryon density ($\rho$)
several times over the nuclear matter density ($\rho_0 = 0.17 {\rm
fm}^{-3}$). However, 
 realistic studies of the high 
density matter based on the
first principle lattice QCD simulation are 
not available yet due to technical difficulties \cite{LQCD}.
In this situation,
any alternative attempts are
welcome to unravel the nature of high density matter.
In particular, how the nuclear matter composed of
 nucleons (which are by themselves  composite three-quark objects)
dissolve into quark matter is an interesting question to
be studied.
 From the experimental and observational point of view,
such transition may occur in high-energy
heavy ion collisions \cite{refQGP} and in the central core of 
neutron stars \cite{Baym95}.

In this Letter, we propose a  molecular dynamics (MD) simulation 
 \cite{refQMD,Mar98PRC}
 of a system composed of many constituent quarks \cite{JAERI99}.
 As a first attempt, we carry out MD simulation 
 for quarks with  SU(3) color
 degrees of freedom. Spin and flavor
 are fixed for simplicity, although there is no
 fundamental problem to include them.
 Time evolution of the spatial and color coordinates of quarks
 are governed by the color
 confining potential, the perturbative 
 gluon-exchange potential and  the meson-exchange potential.
 The confining potential favors the 
 color neutral cluster (nucleon) at low density. However,
 as the baryon density increases, the system 
 undergoes a transition to the deconfined quark matter, since 
  the nucleons start to overlap with
 each other.  Our color MD simulation (CMD) is a natural framework
 to treat such a  percolation transition.
 The meson-exchange potential between quarks,
  which represent the non-perturbative QCD effects, helps
 to prevent the system to collapse. 
 Although techniques are quite different,  
 physical idea behind  CMD with 
 the meson-exchange potential 
 is quite similar in spirit
 with the quark-meson coupling (QMC) model extensively used
 to study the nuclear matter from quarks \cite{QMC}.

We start with a total wave function of the system $\Psi$ as a direct
product of single-particle quark wave-functions. 
The antisymmetrization is neglected at present.
\begin{eqnarray}
\Psi&=&\prod_{i=1}^{3A} \phi_i({\bf r})\chi_i, \\
\phi_i({\bf r})&\equiv&
(\pi L^2)^{-3/4}\exp[-({\bf r}-{\bf R}_i)^2/2L^2-i{\bf P}_i{\bf r}],\\
\chi_i&\equiv&\left(\matrix{
\cos\alpha_i\;e^{-i\beta_i}\;\cos\theta_i \cr
\sin\alpha_i\;e^{+i\beta_i}\;\cos\theta_i \cr
\sin\theta_i\;e^{i\varphi_i}
}\right).
\end{eqnarray}
 Here $A$ is the total baryon number of the system,
 $\phi_i$ is  a Gaussian wave packet centered around ${\bf R}_i$ 
 with momentum ${\bf P}_i$ and  a fixed width $L$.
 $\chi_i$ is a coherent state  in the color SU(3) space
 parametrized by four angles, $\alpha_i, \beta_i, \theta_i$ and $\varphi_i$.
 Although general SU(3) vector has six real parameters,
 the normalization condition $|\chi_i|=1$ and 
 the unphysical global phase reduce
 the number of genuine parameters to four.
 Note that SU(2) spin coherent state parametrized by two angles
 has been used in the MD simulation of the many-nucleon system  with spin 
 \cite{Fel90}.

Time evolution of the system is given by solving the equations of motion
for \{${\bf R}_i$, ${\bf P}_i$, $\alpha_i$, $\beta_i$, $\theta_i$, 
$\varphi_i$\} obtained from the 
 time-dependent variational principle
\begin{eqnarray}
{\partial{\cal L}\over\partial q}
&=&{d\over dt} {\partial{\cal L}\over\partial \dot q},
\end{eqnarray}
together with the  classical Lagrangian, 
\begin{eqnarray}
{\cal L}
&=&\langle\Psi|i\hbar{d\over dt}-\hat H|\Psi\rangle\\
&=&\sum_i[
 -\dot {\bf P}_i{\bf R}_i+\hbar\dot\beta_i\cos2\alpha_i\;\cos^2\theta_i
 -\hbar\dot\varphi_i\sin^2\theta_i]-H,
\end{eqnarray}
where $H = \langle \Psi \mid \hat{H} \mid \Psi \rangle $.  
 The explicit form of the   
  equations of motion reads :
\begin{eqnarray}
\dot {\bf R}_i
&=&{\partial H\over\partial {\bf P}_i}, \ \ \ \ 
\dot {\bf P}_i
=-{\partial H\over\partial {\bf R}_i},\\
\dot\beta_i
&=&-{1\over2\hbar\sin2\alpha_i\;\cos^2\theta_i}{\partial H\over\partial\alpha_i},\\
\dot\theta_i
&=&{1\over2\hbar\sin\theta_i\;\cos\theta_i}{\partial H\over\partial\varphi_i},\\
\dot\alpha_i
&=&{1\over2\hbar\sin2\alpha_i\;\cos^2\theta_i}{\partial H\over\partial\beta_i}
   -{\cos2\alpha_i\over2\hbar\sin2\alpha_i\;\cos^2\theta_i}{\partial H\over\partial\varphi_i},\\
\dot\varphi_i
&=&-{1\over2\hbar\sin\theta_i\;\cos\theta_i}{\partial H\over\partial\theta_i}
  +{\cos2\alpha_i\over2\hbar\sin2\alpha_i\;\cos^2\theta_i}{\partial H\over\partial\alpha_i}.
\end{eqnarray}

As for the color-dependent quark-quark interaction, we employ
 the one-gluon exchange and the linear confining potentials.
 To take into account the essential part of the nuclear force, namely,
 the state independent short range repulsion and the medium range
attraction,
 we include the $\sigma$+$\omega$ meson-exchange 
 potential acting between quarks following ref.\cite{QMC}.
 The total Hamiltonian is written as
\begin{eqnarray}
\hat{H}&=&\sum_i\sqrt{m^2+\hat{{\bf p}_i}^2}+{1 \over 2}
 \sum_{i,j\ne i} \hat{V}_{ij},\\
\hat{V}_{ij}&=&
   -  \sum_{a=1}^8 t_i^a t_j^a V_{\rm C}(\hat{r}_{ij}) 
   +  V_{\rm M}(\hat{r}_{ij}),\\
V_{\rm C}(r)&\equiv& Kr -{\alpha_{\rm s} \over r},\\
V_{\rm M}(r)&\equiv&
  - {g_{\sigma q}^2 \over 4 \pi} {e^{-\mu_{\sigma} r}\over r}
  + {g_{\omega q}^2 \over 4 \pi} {e^{-\mu_{\omega} r}\over r},
\end{eqnarray}
where $t^a = \lambda^a /2$ with 
 $\lambda^a $ being the Gell-Mann matrices,
$V_{\rm C}$ is the confinement and one-gluon exchange terms,
and $V_{\rm M}$ is the meson exchange term \cite{pot}.
 We introduce a smooth infrared cutoff to 
 the confining potential in $V_{\rm C}(r)$ to prevent the 
 long-range interaction beyond the size of the box in which
 we carry out MD simulations.
 We choose the cutoff scale $r_{\rm cut}=3.0$ fm, which is 
 approximately half of the length of the box.
  Typical values of the parameters in the quark model for
 baryons read \cite{pot}, 
$m=350$ MeV (the constituent-quark mass), $\alpha_{\rm s}=1.25$
 (the QCD fine structure constant), 
 $K=0.75$ GeV/fm (the string tension). 
 The meson-quark coupling constants $g_{\sigma (\omega) q}$ are estimated 
 from the meson-nucleon couplings $g_{\sigma (\omega) N}$ 
 using the additive quark picture:
    $g_{\sigma q}= g_{\sigma N}/3 = 3.53$ and 
$g_{\omega q}= g_{\omega N}/3 = 5.85$.
 The meson masses are taken to be
 $\mu_{\omega}=782$ MeV and $\mu_{\sigma}=550$ MeV.

 Some comments are in order here on the evaluation of the
  matrix elements $H = \langle \Psi \mid \hat{H} \mid \Psi \rangle $.

\noindent
(i) We have not taken into account the anti-symmetrization of 
 quarks in the total wave function.
 Because of this, the interaction between quarks in 
 a color-singlet baryon is underestimated by factor 4
 when one takes the matrix element of $t_i^a t_j^a$.
 To correct this, we use effective couplings $K^{\rm eff}=4K$ and
 $\alpha_s^{\rm eff}=4 \alpha_s$ throughout our  CMD simulation.

\noindent
(ii)  $L$ (the size of the quark wave-packet) 
 is chosen to be 0.35 fm (corresponding to 
 the r.m.s. radius of the constituent quark of 0.43 fm). 
 This is consistent with the typical value
 expected from the dynamical breaking of chiral symmetry 
 \cite{vw}. This value is to be used for taking the
 matrix element of the gluonic interaction $V_{\rm C}$.
  On the other hand, the meson-quark coupling is
 intrinsically non-local, since $\sigma$ and $\omega$ have
 their own  quark structure.
  Besides, the meson-exchange interaction between nucleons
 with the nucleon form-factor should be 
 properly reproduced by the superposition of the
 meson-exchange interaction between quarks.
 To take into account these facts, we use $L^{\rm eff} = 0.7$ fm
 (corresponding to the r.m.s. radius of 0.86 fm)  
 in taking the matrix element of $V_{\rm M}$.

\noindent
(iii) $H = \langle \Psi \mid \hat{H} \mid \Psi \rangle $
  generally contains a kinetic energy
 originating from momentum variances of wave packets.
 However, when the width of the wave packet is fixed as a time-independent
 parameter, this kinetic energy is spurious and neglected
 in the present calculation.

 Let us now describe how to simulate   the
 simplest three-quark system, namely the nucleon, in CMD.
  We first search for 
 a three-quark state obeying the  color neutrality condition
 \begin{eqnarray}
 \sum_{i=1}^3\langle\chi_i|\lambda^a|\chi_i\rangle=0
 \hspace{1cm}(a=1,\cdots,8).
 \end{eqnarray}
 This  is satisfied 
 by solving a cooling equation of motion in the color space 
 with a potential proportional to
 $\sum_{i,j\ne i}\sum_{a=1}^8 \langle\chi_i|\lambda^a|\chi_i\rangle
 \langle\chi_j|\lambda^a|\chi_j\rangle$ with random
 initial values of $\chi_i$.
 During this cooling procedure, the spatial coordinates
 of quarks are fixed, e.g. at the three corners of a triangle.

 If we start with three quarks in triangular position
  obtained above  and kick each quark 
  by a same amount of energy 
  keeping the total
  momentum zero, the quarks start to have a breathing motion
  in 2-dimensional plane.
   Since the total color is conserved, 
  the color-neutrality is maintained during this
  time evolution.

 By an initial kick to give the time-averaged kinetic energy of 74 MeV, 
 the total energy  of the nucleon become 1269 MeV.
 Accordingly,  the r.m.s.~radius of the nucleon reads
 0.46 fm in terms of $L$ (which corresponds to the size
 of the quark-core of the nucleon) or 
 0.87 fm in terms of $L^{\rm eff}$ (which corresponds to
 the physical nucleon size for meson-exchange interaction).
  The ``nucleon'' here is certainly a semiclassical object 
 which should be  regarded as a mixture of the
 ground and excited states of three quarks.
  We use a collection of these nucleons as an initial condition 
  for the CMD simulation of many quarks.
 Since the interaction among quarks 
 in matter  will eventually randomize the internal motion of
 quarks in the initial nucleon, 
 the way how we kick the quarks 
 does not matter 
 for the final result.

 Now, let us study  the phase change from
 the  confined hadronic system to 
 the  deconfined quark matter.
 We simulate the infinite matter under the periodic boundary condition
 and see how the system responds to the change of the 
 baryon density as well as to the energy deposition from outside.

 To start with, 
 nucleons constructed as above are randomly distributed
 in a box with the periodic boundary condition.
  At this stage, the total system is in its excited state.
 The minimum energy state of matter is obtained by the frictional
cooling procedure, namely we solve a cooling equation of motion 
with frictional terms. 
 During the cooling, 
 spatial and color motion of quarks in the nucleon are artificially
 frozen, and the following equations are solved:
\begin{eqnarray}
& & \dot {\bf R}_i=
  {1\over 3}\sum_{j\in \{i\}} \left[{\partial H\over \partial {\bf P}_j}
  +\mu_R {\partial H\over \partial {\bf R}_j}\right], \\
& & \dot {\bf P}_i=
  {1\over 3}\sum_{j\in \{i\}} \left[-{\partial H\over \partial {\bf R}_j}
  +\mu_P {\partial H\over \partial {\bf P}_j}\right], \\
& & \dot\alpha_i=\dot\beta_i=\dot\theta_i=\dot\varphi_i=0,
\end{eqnarray}
 where $\mu_R$ and $\mu_P$ are damping coefficients with negative values 
 and $\{i\}$ means a set of three quarks in a nucleon to which $i$ belongs.
  Under this cooling procedure, the system approaches to 
 a stable configuration 
 with minimum energy.
 The system does not collapse due to the repulsive part of the
 meson exchange potential $V_{\rm M}$.

After the system reached its energy-minimum by the cooling,
 internal color and spatial motion of quarks are turned on
 and the normal equation of motion is solved for several tens 
 of fm/c so that the system gets equilibrated.
  To study the excited state of the system,
 extra random motion is also given to the nucleons 
 so that the system has a certain excitation energy.

We judge  the confinement/deconfinement 
 by the following criterion.
If three quarks are within a certain distance $d_{\rm cluster}$
 and are
white with an accuracy $\varepsilon$,   
 these quarks are said to be confined.
 This can be formulated as
\begin{eqnarray}
&&
  \cases{
  \left |{\bf R}_i-{\bf R}_j\right |<d_{\rm cluster}  \ \ (i,j=1,2,3),
  \cr
  \displaystyle
\sum_{a=1}^8\left[\sum_{i=1}^3 \langle\chi_i|\lambda^a|\chi_i\rangle\right]^2 
     <\varepsilon .
  }
\end{eqnarray}
All quarks are checked by this criterion without
 duplications.
 The actual numbers we use are $d_{\rm cluster} = 1 $~fm and
 $\varepsilon = 0.05$.

 Snapshots of matter in equilibrium for different excitation energies 
 per quarks ($E^*$) are displayed in Fig.~1. 
Quarks in the confined states are shown with thin colors and
those in the deconfined state with thick colors.
 As $E^*$ increases, number of deconfined quarks increases as 
 expected. However, 
 some deconfined quarks still form three-quark clusters even
 for large $E^*$.
 This implies that the deconfinement is caused not only by 
 disintegration or percolation  of clusters in the coordinate space but 
also by the color excitation  inside each cluster.

\begin{figure}
\epsfysize=11cm
\centerline{\epsfbox{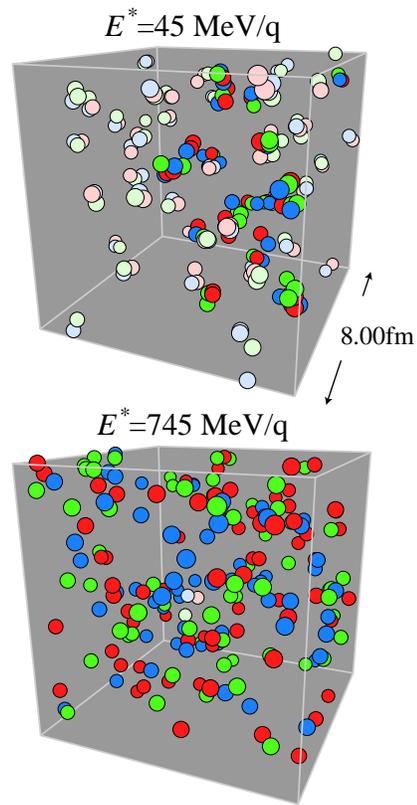}}
\caption{
Perspective of matter with different excitation energies per quark
 $E^*$ at baryon density $0.78\rho_0$.
 Particles with thin (thick) colors indicate quarks in
 the  confined (deconfined) state.}
\end{figure}

Figure 2 shows a ``confined ratio of quarks'', $R \equiv$
 (number of confined quarks)/(total quark number).
 At $E^* =0$, hadronic matter and the quark matter are
 well characterized by $R$ although no sudden
 transition of $R$ between the two phases is observed. 
 For  $E^*> 200$ MeV/q, $R$ is less than 20\% for all densities.

 To study the ``thermal'' property of the system,
 we fit the kinetic energy distribution of quarks
   by the classical Boltzmann distribution.
  Then,  we can
 define an effective temperature $T^*$ for given  $E^*$.
 Note that $T^*$ is not really a physical temperature of the system,
  but is a measure of the averaged kinetic energy per quark.  
 In Fig.~3, plotted is $T^*$ as a function of $E^*$.
 For $E^* > 300$ MeV/q, $T^*$  depends almost linearly on $E^*$ 
 irrespective of baryon density.
 However, for $E^* = 100 \sim 200 $ MeV/q, 
  $T^*$ for low-density matter 
 increases rather slowly as a function of $E^*$.
 In fact, this corresponds exactly to the region where
  the confined ratio of low-density matter changes in Fig.~2.
 This implies that, 
 during the deconfinement process, the energy deposit from outside 
 is consumed to melt the confined clusters (nucleons), which suppresses the 
 effective temperature $T^*$.


\begin{figure}
\epsfysize=5cm
\centerline{\epsfbox{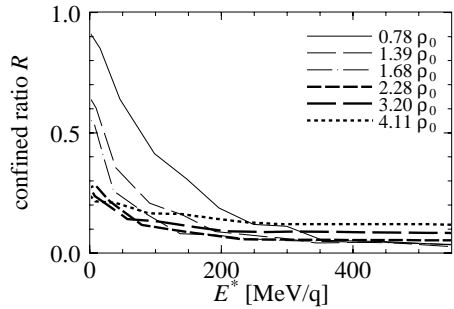}}
\caption{
Confined ratio of quarks as a function of $E^*$ for
 several different baryon densities.}
\end{figure}

\begin{figure}
\epsfysize=5cm
\centerline{\epsfbox{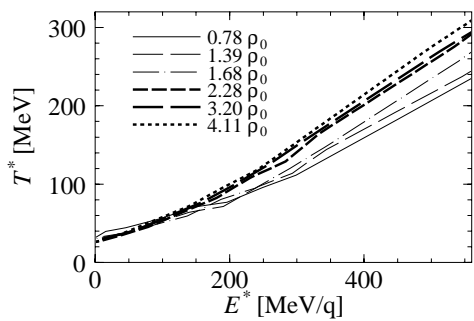}}
\caption{
Baryon density and excitation energy dependence of the effective temperature
extracted from the Boltzmann fit of the kinetic energy distribution.
}
\end{figure}

In summary, we have proposed a color molecular dynamics (CMD) simulation
 of the system with many constituent quarks.
 The system is approximated by  the product of the 
 wave packets with  SU(3) color coherent state.
 Adopting the effective interaction between quarks,
   we study the transition from the nuclear matter to quark matter 
 under the periodic boundary condition. 
 At low baryon density ($\rho$) and low excitation energy ($E^*$),
 the system is in the confined phase where most of the
 quarks are hidden inside the color singlet nucleons.
 However, 
 as  we increase $\rho$ and/or $E^*$,  
 the  partial deconfinement takes place 
 due to the disintegration of color-singlet clusters both in the 
 coordinate space and in the color space.
 This can be seen explicitly by the confined ratio 
 and effective temperature in Fig.2 and Fig.3.

 The results of this paper are still in the qualitative level.
 The refinement of interaction parameters, and the inclusion of
 flavor and spin degrees of freedom as well as 
 anti-quarks  are necessary for more quantitative discussions.
  The use of the antisymmetrized quark wave function 
  is also an important future problem  \cite{refAMD}.
 The medium modification of the constituent-quark mass
 should be also considered in relation to the partial restoration of 
 chiral symmetry.
 In spite of 
  all these reservations, the method proposed in this paper gives  a 
 starting point to study 
   the statistical feature of the hadron-quark transition
 as well as to examine  finite nuclei and the dynamics of  
 heavy-ion collisions. Some preliminary simulation on the latter
 problem has been reported in \cite{JAERI99}.

The authors thank  Y.~Nara, V.~N.~Kondratyev,
 S.~Chikazumi, K.~Niita, S.~Chiba, T.~Kido and A.~Iwamoto
 for
useful suggestions and stimulating discussions.
 T.~H. was partly supported by Grand-in-Aid for
 Scientific Research No. 10874042 of the Japanese
 Ministry of Education, Science and Culture, and by
 Sumitomo Foundation (Grant no. 970248).

\end{document}